\documentclass[article, reprint, superscriptaddress, showpacs, showkeys, prd,14pt]{revtex4-1}

\usepackage{amsmath} 
\usepackage{graphicx} 
\usepackage{verbatim} 
\usepackage{color} 
\usepackage{subfigure} 
\usepackage{hyperref} 
\usepackage{longtable}
\usepackage{dcolumn}
\begin{document}

\title{A generalized dilaton gauge field for the $\rho$ meson mass spectrum in the soft-wall AdS/QCD}

\author{Vlasios Petousis}  \email{vlasios.petousis@cern.ch \\ vpetous@irb.hr} 
\affiliation{Institute Rudjer Boskovic. Bijencka-cesta 54, HR-10000 Zagreb, Croatia}

\date{\today}

\begin{abstract}
Meson spectroscopy within the frame of the soft-wall AdS/QCD, became one of the most interesting topics of particle physics in the last six years. In this work we attempt a generalized parametric analysis of the background dilaton gauge field which previous studied by K.M.Keita and Y.H.Dicko. Aim is that using a positive z-depended dilaton gauge field and setting the appropriate parameters on it, we are able to reproduce the full $\rho$ vector meson mass spectrum. To reinforce the results, we compare the proposed parametric model with the experimental data. This comparison, returns in some cases an error less than $1\%$ and confirms the previous modifications and work which has be done.  
\end{abstract}

\maketitle

\section{Introduction}
Since the 1970's the progress towards to understand the strong interaction has developed surprisingly. Quantum Chromodynamics (QCD) has established as a very successful quantum field theory description of the strong interaction and is by now well tested experimentally. The matter degrees of freedom in QCD consist of quarks transforming in representation of a non-abelian SU(3) gauge theory. Interactions are mediated by gauge bosons, the gluon fields, in the adjoint representation of SU(3). The theory has been shown to be asymptotically free \cite{1},\cite{2}. According to this at large energy scales or equivalent at very short distances the quarks becomes weakly interacting. Contrary at long distances or low energies the force becomes stronger. As result the quarks are confined into bound states, the hadrons. Also in addition the large mass of the quarks, breaks the chiral symmetry, mixing left and right handed quarks. One decade earlier the discovery of QCD in 1960, a radical theory has originated, was the string theory \cite{3},\cite{4},\cite{5},\cite{6},\cite{7}. Found that the hadron spectra contains Regge trajectories, a step further was the realization that these trajectoties can be reproduced by a rotating relativistic string. Those days the four – dimensional string theories lead to non physical modes like tachyons and massless vector particles. So the string theory rejected as a theory which can describe the strong interaction and took another way. String theory today is a potentially promising candidate for the unified theory of all fundamental interactions. The success both of QCD and string theory is enormous but many unsolved problems remains an open issue until today. For the QCD despite the help of the powerful computing (lattice QCD) \cite{8},\cite{9},\cite{10} the low energy mechanism for the confinement and chiral symmetry braking, remains blur. The string theory in that approach can as mentioned before can be a favorable potential candidate but is more than desirable the connection with the experiments. Many open issues and questions it seems that have found a hint of understanding since 1990's. One of the relations between superstring theory and QCD done in 1995, introducing the the concept of D-branes \cite{11},\cite{12},\cite{13}. In 1997 a radical idea about a dual interpretation of D-branes putted forward by Juan Maldacena \cite{14}. In a few words one can say that in Maldacena's duality we can see the matching between a symmetric strongly coupled large N gauge theory and weakly coupled supergravity theory. The gauge theory is a 3+1 dimensional theory which comes out from a basic D3 brane configuration. N number of branes generates an SU(N) gauge theory in the lowest energy limit and to have a supergravity weakly coupled the theory needs a large number of branes. The introduction of QCD – like new theories is necessary to break supersymmetry and remove conformal invariance as well to introduce quark fields. String theory models have inspired phenomenological approaches to QCD gives us the AdS/QCD. More concretely the AdS/CFT correspondence is a duality between the type IIB tring theory defined on the ${AdS_{5}}\times {{S}^{5}}$ space and a  {{\it{N}}={4}}  super Yang-Mills theory with gauge group ${SU{(N_{c})}}$ for large ${N_{c}}$. Little later proposed that the correspondence could be generalized as an a equivalence between a theory defined an ${AdS_{d+1}}\times {L}$ and a conformal field theory living on the flat boundary ${M_{D}}$ of the AdS space \cite{15}. This has provided a new hope in the understanding of strong interaction processes by string inspired approaches. Two main ways followed to achieve this goal. The first so-called top-down approach, for this one starts from string theory and tries to derive a low-energy QCD like theory on ${M_{D}}$ through a compactifications of the  extra dimensions. The second is the bottom-up approach in which one starts from 4D QCD and tries to construct a higher dimensional dual theory \cite{16}. Both approaches must break conformal invariance, since QCD is not a conformal theory \cite{17} and to achieved confinement. The bottom-up approach has two ways to achieved also. The first is the hard-wall model which uses a five dimensional "AdS-slice" with the fifth holographic coordinate z varying up a ${z_{max}}$. Another proposal to break conformal invariance can be done introducing at 5D AdS space, a background dilaton field. The latter model named soft-wall model \cite{18},\cite{19}.
The backbone of this paper organized as follows: in section II presented a sort basic review of the soft-wall model and in section III introduced the proposed dilaton gauge field. In section IV the masses spectrum of $\rho$ vector meson calculated and comes in comparison with the experiment. Finally this paper ends with the conclusions and the appropriate bibliography. 

\section{A review in Soft-Wall AdS/QCD}
To describe the soft-wall AdS/QCD model one can start from QCD and then tries to build its gravity dual theory. 
A good point to start is the Dirac Lagrangian. This Lagrangian has a global $SU{{({{N}_{f}})}_{L}}\times SU{{({{N}_{f}})}_{R}}$ symmetry in a massless limit. The chiral symmetry is then broken by the mass term. Furthermore there is another cause for this that so-called quark condensate and that brakes spontaneously the chiral symmetry. The spectra of mesons in a low-energy QCD are linear and the quarks are confined inside the hadrons. This confinement can be realized in any correct model of low-energy QCD. 
Constructing the holographic dual of QCD one must start with these basic ingredients. Each global symmetry of QCD becomes a local symmetry in the gravity (holographic) side. Thus, in a gravity side, there is one left-handed gauge vector field ${A_{L}}$ for the global $SU{{({{N}_{f}})}_{L}}$ symmetry of QCD and one right-handed gauge vector field ${A_{R}}$ for the global $SU{{({{N}_{f}})}_{R}}$ symmetry of QCD. The spontaneously breaking of the chiral symmetry, achieved by introducing a complex scalar field ${X(z)}$. This complex scalar field belongs to the adjoint representation of the 5D gauge group $SU{{({{N}_{f}})}_{L}}\times SU{{({{N}_{f}})}_{R}}$. One to achieve a linear confinement feature of QCD, simply has to turn on the holographic coordinate dependent dilaton field $\Phi{}$ with the limit $\Phi (z\to \infty )\sim {{(kz)}^{2}}$ where the parameter k sets the meson mass scale. Also the dilaton background is asymptotically quadratic as, $z\to \infty$. According to the general rules of gauge/gravity duality, there are two 5D gauge fields ${(A_{L})}$ and ${(A_{R})}$ which are dual to 4D chiral currents  $J_{L}^{\alpha \mu }=\bar{\psi }_{L}^{\mu }{{\gamma }^{\mu }}{{t}^{\alpha }}{{\psi }_{L}}$ and  $J_{R}^{\alpha \mu }=\bar{\psi }_{R}^{\mu }{{\gamma }^{\mu }}{{t}^{\alpha }}{{\psi }_{R}}$. The quark bilinear operator $\bar{\psi }_{L}^{i}\psi _{R}^{j}$  is also an important 4D operator of chiral symmetry breaking. The energy scale in the 4D theory comes from holographic coordinate z. This coordinate can be defined within the range $0<z<\infty $. The bulk action which describes the mesons sector considering only quadratic parts of the field can be written as:

\begin{equation}
\begin{split}
{{S}_{5}}=\int{{{d}^{5}}x\sqrt{G}{{e}^{-\Phi(z)}}Tr\left[{{\left| DX \right|}^{2}}-m_{x}^{2}{{\left| X \right|}^{2}}
-\frac{1}{4g_{5}^{2}}(F_{L}^{2}+F_{R}^{2})\right]}\label{eq:1}
\end{split}
\end{equation}

where ${{F}_{L}}$ and ${{F}_{R}}$ are the non-Abelian field strength formed from gauge potentials ${(A_{L})}$ and ${(A_{R})}$ respectively and defined in one form as: 
\begin{equation}F_{L,R}^{MN}={{\partial }^{M}}A_{L,R}^{N}-{{\partial }^{N}}A_{L,R}^{M}-i[A_{L,R}^{M},A_{L,R}^{N}]\label{eq:2}\end{equation}
In equation (\ref{eq:1}) the symbol ${D}$ is the Yang-Mills covariant derivative containing the gauge fields $\left( {{A}_{L}},{{A}_{R}} \right)$ and ${ m_{x}^{2}= -3}$. In the integration measure, ${G}$ denotes the determinant of the metric ${(D_{MN})}$, so $\sqrt{G}={{a}^{5}}$ for the ${AdS_{5}}$ space.  
Also the ${D_{M}X}$ can be written as ${{D}_{M}}X={{\partial }_{M}}X-i{{A}_{LM}}X+iX{{A}_{RM}}$  
and the gauge coupling ${g_{5}}$ to be $g_{5}^{2}=\frac{12{{\pi }^{2}}}{{{N}_{c}}}=4{{\pi }^{2}}$ and for QCD, ${{N}_{c}}=3$.
The metric in a 5D AdS geometrical background written as:
\begin{equation}d{{s}^{2}}={{G}_{MN}}d{{x}^{M}}d{{x}^{N}}=a^{2}(z)({{\eta }_{\mu \nu }}d{{x}^{\mu }}d{{x}^{\nu }}-d{{z}^{2}})\label{eq:3}\end{equation}
where ${{\eta }_{\mu \nu }}$ is the 4D Minkowski metric given by:  ${{\eta }_{\mu \nu }=diag(1,-1,-1,-1)}$ and $\alpha (z)$ is the conformal factor (or warped factor) equal to 1/z.
We assumed that the vacuum expectation value (vev), $\upsilon(z)$ of the scalar field have a z-dependent behavior.
Using the equation(\ref{eq:1}) then the equation of motion (EOM) of the vacuum expectation value in axial gauge can written as:
\begin{equation}{{\partial }_{z}}({{a}^{3}}{{e}^{-\Phi }}{{\partial }_{z}}\upsilon )+3{{a}^{5}}{{e}^{-\Phi }}\upsilon =0\label{eq:4}\end{equation}
Hadrons are defined as a normalized modes of the 5D gauge fields and the vector mesons (V) and pseudovector (axial- vector) mesons (A) can be written respectively:  
\begin{equation}V=\frac{{{A}_{L}}+{{A}_{R}}}{2}\label{eq:5}\end{equation}  and  
\begin{equation}A=\frac{{{A}_{L}}-{{A}_{R}}}{2}\label{eq:6}\end{equation}
According to Kaluza-Klein (KK), a dimensional reduction of a 5D vector field, gives us a tower of 4D massive vector fields they so-called KK-modes. These KK-modes can be decomposed by slicing the field in an infinite tower of 4D components.
The vector meson field (V) can be written then: 
\begin{equation}{{V}_{\mu }}(x,z)=\sum\nolimits_{n}{\rho _{\mu }^{(n)}(x)}h_{V}^{n}(z)\label{eq:7}\end{equation} where the z-dependent part satisfies the \cite{20}:
\begin{equation}-{{\partial }_{5}}(a{{e}^{-\Phi }}{{\partial }_{5}}h_{V}^{(n)})=a{{e}^{-\Phi }}{{(M_{V}^{(n)})}^{2}}h_{V}^{(n)}\label{eq:8}\end{equation}  where ${{(M_{V}^{(n)})}^{2}}$ are the masses of vector fields $\rho _{\mu }^{(n)}(x)$. 
The infinite tower of 4D massive vector fields $\rho _{\mu }^{(n)}(x)$ as a result from the KK decomposition of the vector field (V), can be assumed as the vector $\rho$-meson of the low-energy QCD and the equation (\ref{eq:7}) identifies its mass spectrum. Furthermore this equation can be changed by setting the  
$h_{V}^{(n)}={{e}^{\left[ \Phi (z)-\log a(z)/2 \right]}}\chi _{V}^{(n)}$ to the Shr\"odinger form: 
\begin{equation}-\frac{{{d}^{2}}}{d{{z}^{2}}}\chi _{V}^{(n)}+{{V}_{V}}\chi _{V}^{(n)}={{(M_{V}^{(n)})}^{2}}\chi _{V}^{(n)}\label{eq:9}\end{equation} with potential: 
\begin{equation}
\begin{split}{{V}_{V}(z)}=\frac{1}{4}{{[\frac{d}{dz}(\Phi (z)-\log a(z))]}^{2}}\\-\frac{1}{2}[\frac{{{d}^{2}}}{d{{z}^{2}}}(\Phi (z)-\log a(z))]\label{eq:10}
\end{split}
\end{equation}

\section{The parametric modified dilaton gauge field}
In the following analysis a parametric modified version of the soft-wall AdS/QCD model according the dilaton gauge field presented. The model has been supported at \cite{21} and modified further by others \cite{22},\cite{23},\cite{24},\cite{25},\cite{26}. About the interesting regions in AdS geometry one can say that a dilaton gauge field can be freely between the IR region with $\Phi \sim {{z}^{2}}$ which is just the requirement for the Regge behavior and the UV with $\Phi \sim {{z}}$. The proposed parametric dilaton field actually is can be written as: \begin{equation}\Phi (z)={{(kz)}^{2}}+f(z)\label{eq:11}\end{equation}
Where the f(z) is the proposed general parametric function which has the form:
\begin{equation}f(z)=\lambda z+\mu \log (z)\label{eq:12}\end{equation}
This parametric function relation, of the dilaton gauge field, has three parameters, k and $\lambda$ which both real numbers and have units at the MeV regime and the $\mu$ which is an integer number.  
Another step is to specify the appropriate parameters on it. Here there are some conditions. More specific we can declare that must: $k>0$, $\lambda \ge 0$ and $\mu \le 0$. Also that introduces a restriction relation between the parameters at the equation (\ref{eq:12}). These must obey the following relation:
\begin{equation}\lambda \ge -\mu \log (\sqrt[z]{z})\label{eq:13}\end{equation}
and that because the dilaton gauge field at the equation (\ref{eq:11}) must be able to reproduce the correct Regge trajectories behavior. The advantage of the positive z-dependent term, which according to \cite{19} is able to avoid massless modes for the vector sector, is that the sign of the exponential profile which defining the wall, should be positive. Also one to define the meson mass spectra needs an additional specification for the warped factor. For our analysis the proposed type of this factor is the following simple: 
\begin{equation}a(z)=\frac{1}{z}\label{eq:14}\end{equation} 

\section{The  $\rho$ meson mass spectrum}
Our analysis focus on the $\rho$ vector meson which is the most represented meson for that purpose.
To simplify our analysis we select to analyze the $\rho$ vector meson mass spectrum using two cases of parameterizations. 
\subsection{The case of $\lambda =k$ and $\mu =0$}
Then the equation (\ref{eq:11}) together with the (\ref{eq:12}) give us the simple dilaton gauge field:
\begin{equation}\Phi (z)={{k}^{2}}({{z}^{2}}+\frac{1}{k}z)\label{eq:15}\end{equation} 
For the first set using the equation (\ref{eq:10}) one can easy evaluate the Shr\"odinger-like representation and for the vector potential can have:
\begin{equation}{{V}_{V}}(z)={{k}^{4}}{{z}^{2}}+{{k}^{3}}z+\frac{{{k}^{2}}}{4}+\frac{1}{2z}\left[ {{k}^{2}}+\frac{3}{2z} \right]\label{eq:16}\end{equation}      
To get the $\rho$ meson mass from equation (\ref{eq:9}) one using the potential given at (\ref{eq:13}) make use of boundary conditions ${{\chi }_{n}}(z\to 0)=0$ and ${{\partial }_{z}}{{\chi }_{n}}(z\to \infty )=0$ and the value for $\lambda$ = k = 402 MeV. One doing this calculations can construct the theoretical masses of the $\rho$ meson mass spectra. This masses given in table I, the experimental values for comparison taken from \cite{23}. 

\begingroup
\squeezetable
\begin{table*}
\caption{The theoretical and experimental values of the $\rho$ meson. The theoretical values are the results of using the first set of parameters to the modified background field. The average error is 5.43\%} 
\begin{tabular}{c c c c c c c c}
\hline\hline
$\rho$  & 0 & 1 & 2 & 3 & 4 & 5 & 6 \\ [0.5ex]
\hline
${{m}_{th}(MeV)}$ & 984 & 1296 & 1543 & 1754 & 1941 & 2139 & 2257\\ 
${{m}_{exp}(MeV)}$ & 775.5$\pm $1 & 1282$\pm$37 & 1465 $\pm$25 & 1720$\pm$20 & 1909$\pm$30 & 2149$\pm$17& 2265$\pm$40\\
Error &  $27\%$ &  $1.1\%$ &  $5.3\%$ &  $2 \%$ &  $1.7\%$ & $0.5\%$ &  $0.4\%$\\[1ex]
\hline\hline
\end{tabular}
\label{table:I}
\end{table*}
\endgroup

\subsection{The case of $\lambda =k$ and $\mu =-1$}
Doing this analysis then the equation (\ref{eq:11}) together with the (\ref{eq:12}) give us the following dilaton gauge field:
\begin{equation}\Phi (z)={{k}^{2}}({{z}^{2}}+\frac{1}{k}z)-\log (z)\label{eq:17}\end{equation}
If we do the same analysis as at the previous case and choose the value $\lambda$ = k = 421 MeV, then we can see that the results our calculations fit  better  for the experimental masses of the $\rho$ vector meson. These results are shown in Table \ref{table:II}. 

\begingroup
\squeezetable
\begin{table*}
\caption{The theoretical and experimental values of the $\rho$ meson. The theoretical values are the results of using the second case of parameters to the modified background field.The average error is 1.96\%} 
\centering 
\begin{tabular}{c c c c c c c c}
\hline\hline
$\rho$  & 0 & 1 & 2 & 3 & 4 & 5 & 6 \\ [0.5ex]
\hline
${{m}_{th}(MeV)}$ & 770 & 1186 & 1483 & 1725 & 1936 & 2124 & 2296\\ 
${{m}_{exp}(MeV)}$ & 775.5$\pm $1 & 1282$\pm$37 & 1465 $\pm$25 & 1720$\pm$20 & 1909$\pm$30 & 2149$\pm$17& 2265$\pm$40\\
Error &  $0.7\%$ &  $7.5\%$ &  $1.2\%$ &  $0.3\%$ &  $1.4\%$ &  $1.2\%$ &  $1.4\%$\\[1ex]
\hline\hline
\end{tabular}
\label{table:II}
\end{table*}
\endgroup

\section{Conclusions}
In this work we studied in the frame of the soft-wall AdS/QCD model, the behavior of the $\rho$ meson mass spectra using a parametric modification of the gauge dilaton background. When this parameters applied, it seems that an improved progress can be achieved. The errors as one can see in Table \ref{table:I} and Table \ref{table:II} are almost within $10\%$. Comparing the two cases one can see very easy that the second case, represents more accurate the experimental values and specially for the $\rho$(770). We can claim that this parametric modification at the gauge dilaton background, indicates that the model has real behavior, is comparable with the experimental data and then we can considered as a general dilaton gauge field for the $\rho$ meson. Furthermore the positive form of the dilaton gauge field satisfies the constrain for having the correct Regge trajectories. This is proof that we don't have massless modes in a vector sector and comes as a second confirmation of the first modification which proposed by K.M.Keita and Y.H.Dicko. 
  
\section{Acknowledgments}
I am grateful to Elias Kiritsis and Johanna Erdmenger for the useful papers which I used as an additional guide for this work and also the K.M.Keita and Y.H.Dicko for their useful papers.

\end{document}